\documentclass[a4paper,11pt]{article}

\usepackage[T1]{fontenc}
\usepackage{amssymb,amsmath,xcolor,slashed,hyperref,comment}
\usepackage[nosort]{cite}

\newcommand{\be}{\begin{eqnarray}}
\newcommand{\ee}{\end{eqnarray}}

\allowdisplaybreaks

\csname @addtoreset\endcsname{equation}{section}
\usepackage{float}
\allowdisplaybreaks

\def\Re{\mathop{\rm Re}\nolimits}

\def\rme{{\rm e}}
\def\rmi{{\rm i}}

\newcommand{\U}{\mathop{\rm {}U}}
\newcommand{\bbox}{\lower.2ex\hbox{$\Box$}}

\newcommand{\jb}{\bar{\jmath}}
\newcommand{\tD}{{\text D}}
\newcommand{\tT}{{\text T}}
\newcommand{\hc}{{\rm h.c.}}
\newcommand{\poinc}{\boxdot}

\begin{document}

\begin{titlepage}
	\thispagestyle{empty}
	\begin{flushright}

		\hfill{DFPD-2017/TH/13}
	\end{flushright}

	\vspace{35pt}

	\begin{center}
	    { \Large{\bf Fayet--Iliopoulos terms in supergravity without gauged R-symmetry}}

		\vspace{50pt}

		{Niccol\`o~Cribiori$^{1,2,3}$, Fotis~Farakos$^{3}$, Magnus~Tournoy$^{3,4}$ and Antoine~Van~Proeyen$^{3}$ }

		\vspace{25pt}

		{
			$^1$ {\it  Dipartimento di Fisica  e Astronomia ``Galileo Galilei''\\
			Universit\`a di Padova, Via Marzolo 8, 35131 Padova, Italy}

		\vspace{15pt}

			$^2${\it   INFN, Sezione di Padova \\
			Via Marzolo 8, 35131 Padova, Italy}

		\vspace{15pt}

			$^3${\it   KU Leuven, Institute for Theoretical Physics, \\
			Celestijnenlaan 200D, B-3001 Leuven, Belgium}

		\vspace{15pt}

			$^4${\it   Institute for Theoretical Physics Amsterdam, 					Delta Institute for Theoretical Physics, \\
			University of Amsterdam, Science Park 904, 1098 XH 							Amsterdam, The Netherlands}

		}
		\vspace{40pt}

		{ABSTRACT}
	\end{center}

	\vspace{10pt}

We construct a supergravity-Maxwell theory with a novel embedding of the Fayet--Iliopoulos D-term, leading to spontaneous supersymmetry breaking.
The gauging of the R-symmetry is not required and a gravitino mass is allowed for a generic vacuum.
When matter couplings are introduced, an uplift through a positive definite contribution to the scalar potential is obtained.
We observe a notable similarity to the $\overline{D3}$ uplift constructions and we give a natural description in terms of constrained multiplets.

\vfill

\hrule width 3.cm
{\footnotesize \noindent e-mails: niccolo.cribiori@pd.infn.it, fotios.farakos@kuleuven.be, magnus.tournoy@kuleuven.be,\\
antoine.vanproeyen@fys.kuleuven.be }

\end{titlepage}
\addtocounter{page}{1}
\baselineskip 6 mm

%\tableofcontents

\section{Introduction}
\label{sec:introduction}

Supersymmetry breaking is commonly sourced by F-terms or by D-terms.
The case of F-term breaking has been studied extensively. One of its issues is related to the existence of scalar modes,
which in general have to be stabilized in the vacuum, in order to obtain de Sitter backgrounds \cite{Kallosh:2014oja}.
D-term supersymmetry breaking, on the other hand, does not essentially involve scalar fields,
as it is constructed from a vector multiplet.
Its prototype example in supergravity is represented by the embedding of the Fayet--Iliopoulos (FI) term \cite{Fayet:1974jb}.

The only known method to couple a pure FI term to supergravity \cite{Freedman:1976uk,Barbieri:1982ac} is related to the gauging of the R-symmetry.
It allows to study D-string solutions, inflation and no-scale models in supergravity \cite{Binetruy:1996xj,Dvali:2003zh,Binetruy:2004hh,DallAgata:2013jtw,Aldabergenov:2017hvp}.
As pointed out in \cite{Komargodski:2009pc},
models of pure D-term supersymmetry breaking with FI terms cannot arise directly from any UV-complete theory of gravity,
as they imply the existence of a global abelian symmetry in the UV regime.
Even though caveats can exist to the applicability of \cite{Komargodski:2009pc},
it is acknowledged that D-term breaking \`a la Fayet--Iliopoulos has either to be excluded,
or to be treated as an effective description of some more fundamental underlying mechanism.

In this work we present a new supersymmetric coupling, which includes only a Maxwell vector multiplet and leads to D-term supersymmetry breaking within supergravity.
The model contains non-linearities similar to supersymmetric Born--Infeld actions \cite{Cecotti:1986gb}.
In sharp contrast
to the case of the D-term breaking arising from a FI term with R-symmetry gauging,
the model we are introducing allows for a non-vanishing gravitino mass term on a generic vacuum.
As it is going to be clarified in the following,
another distinctive property of the new term is that it can be consistent if and only if supersymmetry is broken by - but not only from - the auxiliary field of the abelian vector multiplet.
As far as we know, this is the first example of D-term breaking in supergravity that has these properties and that is not introducing higher derivatives in the bosonic sector (for example, models with higher derivatives were presented in \cite{Farakos:2013zsa,Fujimori:2017kyi}).
In addition, as the model we are presenting does not require the gauging of the R-symmetry, the results of \cite{Komargodski:2009pc} do not apply.

To illustrate the properties of the new coupling, in the following
we first construct a pure model that contains only the abelian gauge multiplet coupled to supergravity
and subsequently we couple it to matter chiral multiplets.
We find then that the generic impact of the new term is to uplift the vacuum energy.
In particular, if we couple to a chiral model with a K\"ahler potential $K$,
we obtain\footnote{We adopt the conventions of \cite{Freedman:2012zz}.}
\begin{equation}
\label{Vintro}
{\cal V} = {\cal V}_{\substack{\text{Standard} \\\text{SUGRA}}} + \frac{\xi^2}{2}\,\rme^{2K/3}\, ,
\end{equation}
where $\xi$ is a real constant and the contribution to the scalar potential \eqref{Vintro} proportional to $\xi^2$ comes from the new term we are introducing.
Notice how this contribution is reminiscent of the uplift arising from an $\overline{D3}$ brane, which can be also described with nilpotent multiplets \cite{Ferrara:2014kva,Kallosh:2014wsa,Bergshoeff:2015jxa}, even though our construction has linearly-realized supersymmetry off-shell.

This work is organized as follows.
In section \ref{sec:freedman_model} we discuss the model of Freedman and we illustrate the associated gauging of the R-symmetry.
In section \ref{sec:new_d_term} we present the new D-term model together with its properties.
In section \ref{sec:matt_coup} we couple the model to matter and show how the uplift is produced.
In section \ref{sec:d_term_nilpotent} we recast this new model into the language of nilpotent multiplets.
In section \ref{ss:concl} we summarize our conclusions and give some comments. In the appendix A we formulate a specific parametrization for the constrained superconformal multiplets, which is useful for section~\ref{sec:d_term_nilpotent},
and in the appendix B we give details on the electric-magnetic duality of the new D-term.

% section introduction (end)

\section{Freedman model and gauged R-symmetry}
\label{sec:freedman_model}

The prototype example of D-term breaking in $\mathcal{N}=1$ supergravity is the model introduced in \cite{Freedman:1976uk}.
In this section we are going to discuss this model in some detail, in order to identify the differences with our new model in the next section.
The model in \cite{Freedman:1976uk} describes the couplings of a gravitational spin-$(2,\frac32)$ multiplet interacting with an abelian spin-$(1,\frac12)$ vector multiplet. The vector field couples both to the gravitino and the spin-$\frac12$ fermion and gauges a chiral symmetry acting on them. This symmetry can be identified with the R-symmetry, as it rotates the fermions of the theory.

In order to reproduce this model within the language of tensor calculus, we introduce a real multiplet $V = \{V,\,\zeta,\,{\mathcal{H}},\, v_\mu,\, \lambda, {\text{D}}\}$ \footnote{In our notation, each multiplet is going to be identified with its lowest component. Higher components are going to be denoted with subscripts, $e.g.$ $(V)_{\zeta}=\zeta$.}, with vanishing Weyl ($w$) and chiral $(c)$ weights, gauged by a chiral multiplet ${\cal B}$
\begin{equation}
\label{Btransform}
V \ \rightarrow\ V + \rmi {\cal B} - \rmi \overline{\cal B} \,,
\end{equation}
and a compensator chiral multiplet $\phi^0=\{\phi^0,\,\Omega^0,\,F^0\}$, with $(w,c)=(1,1)$.
The compensator multiplet transforms under (\ref{Btransform}) as
\begin{equation}
  \phi ^0\ \rightarrow\ \phi ^0 \exp \left( -\frac{2}{3}\kappa ^2\rmi\xi {\cal B} \right) \,,
 \label{phiBtransform}
\end{equation}
where $\xi$ is a real constant and the normalizations are the ones adopted in \cite{Freedman:2012zz}. In the Wess-Zumino gauge, the first components of $V$ are put to zero: $v=\zeta ={\cal H}=0$, and only the gauge multiplet $\{v_\mu,\, \lambda,\, {\text{D}}\}$ is non-vanishing. The only effective part of ${\cal B}$ is the real part of the lowest component: $\theta = -2\Re {\cal B}$, acting as
\begin{equation}
  \delta (\theta ) \phi ^0= \frac1{3}\kappa ^2\rmi\,\theta\, \xi\, \phi ^0\,,\qquad  \delta (\theta )v_\mu =\partial _\mu \theta \,.
 \label{delthetaphi0}
\end{equation}

The Lagrangian in the superconformal setup reads
\begin{equation}
\label{LVB}
{\cal L}
= -3 \Big{[} \phi^0 \overline{\phi} ^0 \Big{]}_D
-\frac{1}{4} \left[\overline\lambda P_L\lambda\right]_F \, .
\end{equation}
In order to obtain Poincar\'e supergravity in the Einstein frame, part of the superconformal symmetries have to be gauge-fixed.
In particular, in this setup one proceeds to the Poincar\'e supergravity by setting $\phi^0=\kappa ^{-1}$.
This mixes the symmetry  (\ref{delthetaphi0}) with the conformal $\U(1)$ symmetry, acting e.g. as
\begin{equation}
  \delta (\lambda _T)\phi ^0 = \rmi \lambda _T\phi ^0,\, \quad \delta (\lambda _T)F ^0=-2\rmi\lambda _TF ^0,\,\quad  \delta (\lambda _T)\psi_\mu = \frac{3}{2}\rmi\gamma _*\lambda _T\psi_\mu \,,\quad
  \delta (\lambda _T)A_\mu =\partial _\mu \lambda _T\,,
 \label{delT}
\end{equation}
keeping after the gauge fixing $\lambda _T=-\frac{1}{3}\kappa ^2\xi \theta $.

Once the conformal symmetry in \eqref{LVB} is gauge fixed and the auxiliary fields are integrated out, supersymmetry is broken by a non-vanishing vacuum expectation value
\begin{equation}
\langle\text{D}\rangle=\xi
\end{equation}
and the goldstino can be identified with the gaugino $\lambda$.
The component form Lagrangian in the gauge in which the goldstino is set to zero is
\begin{equation}
\begin{aligned}
e^{-1}\mathcal{L}
=& \frac1{2\kappa ^2}\left(R(\omega(e,\psi))-\overline{\psi}_\mu \gamma^{\mu\nu\rho}\left(\partial_\nu+\frac14 {\omega_\nu}^{ab}(e,\psi)\gamma_{ab}+\frac12\rmi\kappa ^2\xi\, v_\nu\gamma_*\right)\psi_\rho\right)\\
&-\frac14 F_{\mu\nu}F^{\mu\nu}
+\frac{3}{\kappa ^2} \left(A_a + \frac{1}{3}\kappa ^2 v_a\xi \right)\left(A^a + \frac{1}{3}\kappa ^2 v^a\xi \right) -3 F^0\overline F^0+\frac12 \text{D}^2-\xi \text{D}\, ,
\end{aligned}
\label{LwithauxFreedman}
\end{equation}
where $F_{\mu \nu }= 2\partial _{[\mu }v_{\nu ]}$ and $F^0$, $A_a$ are the two auxiliary fields of supergravity in the old-minimal formulation.
As anticipated, in the covariant derivative of the gravitino the vector $v_\mu$ has the role of gauge connection for the local R-symmetry.
It is important to notice that there exists a smooth limit for $\xi \rightarrow 0$ in which supersymmetry is restored.

By integrating out the auxiliary degrees of freedom, the following expression for the on-shell Lagrangian can eventually be obtained
\begin{equation}
\label{FRonshell}
\begin{aligned}
e^{-1}\mathcal{L}
=& \frac1{2\kappa ^2}\left(R(\omega(e,\psi))-\overline{\psi}_\mu \gamma^{\mu\nu\rho}\left(\partial_\nu+\frac14 {\omega_\nu}^{ab}(e,\psi)\gamma_{ab}+\frac12\rmi \kappa ^2\xi\, v_\nu\gamma_*\right)\psi_\rho\right)
\\
&-\frac14 F_{\mu\nu}F^{\mu\nu}-\frac12\xi^2 .
\end{aligned}
\end{equation}
The positive contribution to the vacuum energy given by the FI term can be recognized.
Extensions of this setup with chiral superfields can be found in \cite{Binetruy:2004hh,Villadoro:2005yq}.

Before ending this section, it is instructive to review why the gauged R-symmetry was essential for the local supersymmetric completion of the FI term in \cite{Freedman:1976uk}, using the Noether method. Recall first of all that, under rigid supersymmetry, D transforms to a total derivative. In supergravity, however, when considering the supersymmetry variation of
\begin{equation}
\label{LFI}
\mathcal{L}_{\rm FI} = -e\,\xi\,\text D,
\end{equation}
we have to take into account that $\tD$ does not transform as a total derivative:
 \begin{equation}
\delta \text D = \frac{\rm i}{2}\overline\epsilon\gamma _*\gamma ^\mu {\cal D} _\mu \lambda  \qquad \rightarrow \qquad \delta \mathcal{L}_{\rm FI} =-\frac{\rm i}{2}e \xi\, \overline \epsilon \gamma_*\slashed{\cal D}\lambda +\ldots \, ,
\label{dDdD}
\end{equation}
where $\epsilon$ is the local supersymmetry parameter. In order to complete its variation in (\ref{LFI}) to a total derivative, one needs at least a gravitino-goldstino mixing term
\begin{equation}
 \label{LMIX}
\mathcal{L}_{\rm mix}=\frac{\rm i}{2}e\xi\,\overline\psi_\mu\gamma^\mu\gamma_* \lambda\,.
\end{equation}
The variation of $\lambda$ to $\tD\epsilon $ in the same term cancels also the contributions $-\frac14 e\,\xi\, \overline \epsilon \gamma^\mu\psi_\mu \text D$, which come from the variation of the vielbein and of $\text D$ in \eqref{LFI}.

The remaining interesting contribution in $\delta \mathcal{L}_{\rm mix}$ is the one from $\delta \lambda $ containing the field-strength $F_{\mu\nu}$, which together with the term in the variation  (\ref{dDdD}) where
\begin{equation}
 {\cal D}_\mu \lambda = \ldots -\frac14 \gamma \cdot F\psi _\mu \,,
 \label{covariantquantities}
\end{equation}
produces
\begin{equation}
\label{Ncrucial}
\delta \left(\mathcal{L}_{\rm FI} +\mathcal{L}_{\rm mix}\right)= e   \frac{\rmi}{4}\,\xi\, \overline \psi_\mu \gamma^{\mu\nu\rho}\gamma_*\epsilon\,\, F_{\nu\rho}+\ldots \,.
\end{equation}
In order to eliminate it, in \cite{Freedman:1976uk} a new term was added, modifying the covariant derivative of the gravitino with a contribution proportional to $v_\mu$. This is the last term on the first line of (\ref{LwithauxFreedman}). To add the vector to the covariant derivative, one has to declare $v_\mu $ as a gauge field for the R-symmetry. This, in turn, also modifies the covariant derivative of the gravitino supersymmetry transformation in the same way. As a consequence, the variation of the gravitino kinetic energy contains the curvature of these covariant derivatives, canceling (\ref{Ncrucial}). The variation of \eqref{LFI} is therefore completely eliminated by adding a gravitino-goldstino mixing term and by promoting the vector $v_\mu$ to be the gauge field of the R-symmetry rotating the gravitino and the other fermions.

In the next section we are going to show at which point of this procedure a difference appears in the construction of the model of Freedman and of the one we are proposing and, in turn, how we avoid the gauged R-symmetry that was necessary in \cite{Freedman:1976uk}.

% section freedman_model (end)

\section{A new D-term}
\label{sec:new_d_term}

In this section we are going to discuss the properties of the new D-term model when it is coupled to supergravity.
The pure abelian vector multiplet sector is always going to be described by the Lagrangian
\begin{equation}
\label{newDterm}
{\cal L}_\text{NEW} =
-\frac{1}{4}\left[ \overline \lambda P_L \lambda \right]_F
- \kappa ^2 \, \left[ \xi \, \phi^0\overline{\phi^0}\frac{w^2 \overline{w}^2 }{\text{T}(\overline{w}^2)\overline{\text{T}}(w^2)} \, \, (V)_{\text D}\right]_D,
\end{equation}
where $\xi$ is a real constant, $w^2$ is the chiral multiplet of (Weyl,chiral) weight~(1,1):
\begin{equation}
  w^2 = \frac{\overline \lambda P_L \lambda }{(\phi^0)^2}\,,\qquad \overline{w}^2 = \frac{\overline \lambda P_R \lambda }{(\overline{\phi^0})^2}\,,
 \label{defw2}
\end{equation}
$\text{T}$ is the operator defined in \cite{Kugo:1983mv,Ferrara:2016een}, which defines a chiral multiplet, and $\overline{\text{T}}$ defines an antichiral multiplet.\footnote{In the superspace formalism of \cite{Wess:1992cp}
the supergravity Lagrangian \eqref{newDterm} has the form
\[
{\cal L}_\text{NEW} = \frac14 \left( \int d^2 \Theta \, 2 {\cal E} \,  {\cal W}^2(V) + \text{h.c.} \right) + 8 \int d^4 \theta \, E \, \xi \frac{{\cal W}^2 \overline{\cal W}^2 }{{\cal D}^2 {\cal W}^2  \overline{{\cal D}}^2 \overline{{\cal W}}^2 } {\cal D}^\alpha {\cal W}_\alpha.
\]}
The chiral multiplet $\text{T}\left(\overline w^2\right)$ has Weyl and chiral weights $(2,2)$, while the multiplet $(V)_{\text D}$ is a real linear multiplet, which has Weyl and chiral weights $(2,0)$. It is the conformal version of the multiplet $K$ defined in \cite{VanProeyen:1979ks}, with components
\begin{equation}
 (V)_{\text D}\equiv \text D= \left\{{\text D},\, \slashed{\cal D}\lambda ,\,0,\, {\cal D}^b \widehat{F}_{ab},\, -\slashed{\cal D}\slashed{\cal D}\lambda,-\bbox^C {\text D}\right\}\,,
 \label{Dmultiplet}
\end{equation}
where the definition of $\slashed{\mathcal{D}}\lambda$ and $\widehat{F}_{ab}$ can be found in \cite[(17.1)]{Freedman:2012zz}.
It has been shown in \cite{VanProeyen:1979ks} that also the first term in (\ref{newDterm}) can be written in terms of this multiplet as
\begin{equation}
  -\frac14 \left[ \overline \lambda P_L \lambda \right]_F=\left[V \,{\text D}\right]_D\,.
 \label{rewritekin}
\end{equation}

The gauge invariance of the Lagrangian \eqref{newDterm} is manifest.
Using the fermionic nature of $\lambda $ and taking into account that
\begin{equation}
 \left( \overline \lambda P_L\lambda \right)_F = 2\overline \lambda P_L\slashed{\cal D}\lambda +\widehat{F}^-\cdot \widehat{F}^- - \text{D}^2\, ,
 \label{lambda2Fcomp}
\end{equation}
one can write (\ref{newDterm}) as
\begin{equation}
{\cal L}_\text{NEW} =-\frac{1}{4}\left[ \overline \lambda P_L \lambda \right]_F
- \kappa ^2\xi \left[\frac{(\phi^0 \overline{\phi^0})^3\, w^2\, \overline{w}^2}{\tD^2_+\tD^2_-} \,{\text D}\right]_D\,,
 \label{newDtermsimplified}
\end{equation}
where
\begin{equation}
    \tD^2_+ = \tD^2 - \widehat{F}^+\cdot \widehat{F}^+\,,\qquad \tD^2_- = \tD^2 - \widehat{F}^-\cdot \widehat{F}^-\,.
 \label{deftD2+-}
\end{equation}
These factors appear in denominators and should thus be non-vanishing. If the vacuum expectation value of $\text{D}$ is vanishing, then
the new $\xi$-term becomes ill defined.
This term has a structure similar to the models of \cite{Cecotti:1986gb}, which make use of the same multiplets, but, as we are going to show, in the bosonic sector it produces a linear term in the auxiliary field D.
The component expansion of \eqref{newDterm} is indeed
\begin{equation}
\label{DDFFDD}
e^{-1} {\cal L}_\text{NEW} = -\frac14 F_{\mu\nu}F^{\mu\nu}
- \frac12 \overline \lambda \slashed{\mathcal{D}} \lambda
+ \frac12 \text{D}^2 - \kappa ^2 \xi \phi^0 \overline{\phi^0}\, \text{D}
+ \text{fermionic interactions} \, .
\end{equation}
Notice that if one integrates out the D auxiliary field from \eqref{DDFFDD} there is a contribution to the pure bosonic sector of the form
\begin{equation}
{\cal L}_\xi = - \frac12 \xi^2 \left( \kappa ^2\phi^0 \overline{\phi^0} \right)^2\,,\qquad  \left.{\cal L}_\xi\right|_\poinc = - \frac12 \xi^2 \,,
\end{equation}
where the last expression uses the Poincar\'{e} gauge $\phi ^0=\kappa ^{-1}$.

Let us now turn to the fermionic interactions arising from the couplings proportional to $\xi$ in \eqref{newDterm}, which are highly non-linear. To obtain these, note that the second term of (\ref{newDterm})
contains a product of the  linear multiplet (\ref{Dmultiplet}) and a real multiplet
\begin{equation}
  \mathcal{R}= \frac{\phi^0\,w^2}{\text{T}\left(\overline w^2\right)}\cdot \frac{\overline{\phi^0}\,\overline{w}^2}{\overline{\text{T}}\left( w^2\right)}
\,,\qquad {\rm i.e.}\quad {\cal L}_\text{NEW} =-\frac{1}{4}\left[ \overline \lambda P_L \lambda \right]_F
- \kappa ^2\xi \left[\mathcal{R}\, {\tD}\right]_D \,,
 \label{Rmultiplet}
\end{equation}
where $\mathcal{R}$ is built from chiral and antichiral multiplets. Such an action simplifies \cite{deWit:1981fh} and for our case the terms that can have at most quadratic terms in the fermions are
\begin{equation}
 2\,e^{-1} \left[\mathcal{R}\, {\tD}\right]_D =  (\mathcal{R})_\tD\, \tD -\overline{(\mathcal{R})}_{\lambda}\slashed{\cal D}\lambda-\frac12 \rmi \tD \overline \psi \cdot \gamma \gamma _* (\mathcal{R})_\lambda -D^bF_{ab}(\mathcal{R})_{v}^a
  +\ldots   \,,
 \label{prodVL}
\end{equation}
To find the relevant components, note that  (\ref{Rmultiplet}) contains three chiral multiplets. The first is the compensating multiplet $\{\phi ^0,\, P_L\Omega ^0, F^0\}$. The second is the multiplet $\overline \lambda P_L\lambda $, whose F-component is  (\ref{lambda2Fcomp}) and its fermionic component is
\begin{equation}
P_L\left(\overline \lambda P_L\lambda \right)_\chi  = \sqrt{2}P_L\left(-\frac{1}{2}\gamma \cdot \widehat{F}+\rmi \tD\right)\lambda \,.
 \label{defLambda}
\end{equation}
The third chiral multiplet is the multiplet $\text{T}\left(\overline w^2\right)$, whose lowest component is
\begin{equation}
\begin{aligned}
  \text{T}\left(\overline w^2\right)&= (\overline{\phi ^0})^{-2}\left(2\overline \lambda P_R\slashed{\cal D}\lambda+\widehat{F}^+\cdot \widehat{F}^+ - \text{D}^2\right)\\
  &-2(\overline{\phi ^0})^{-3}\left[\sqrt{2}\overline{\Omega }^0P_R\left(\frac{1}{2}\gamma \cdot {F}+\rmi \tD\right)\lambda
   +\overline{F}^0\overline \lambda P_R\lambda \right]+\text{4-fermions} \,.
 \end{aligned}\label{lowestT}
\end{equation}
Its fermion component is for the part linear in fermions
\begin{equation}
 P_L \left(\text{T}\left(\overline w^2\right)\right)_\chi = -\sqrt{2}P_L\slashed{\cal D}\left[(\overline{\phi^0} )^{-2}\left(\frac{1}{2}\gamma \cdot {F}+\rmi \tD\right)\lambda  \right]+\ldots \,
 \label{Tchi}
\end{equation}
and the $F$ component has no pure bosonic part. The components of $\mathcal{R}$ can be found then from the tensor calculus rules. They take a simple form in terms of the composite fermion
\begin{equation}
\label{calG}
{\cal G} = 2\overline{\phi^0} \frac{ \left( \frac12\gamma \cdot \widehat{F}^- - {\rm i} \text{D} \right) }{\text{D}^2_-} P_L \lambda
+ 2\phi^0 \frac{ \left(\frac12\gamma \cdot \widehat{F}^+ +  {\rm i} \text{D} \right) }{\text{D}^2_+} P_R \lambda \,,
\end{equation}
whose left component has weights $(w,c)=(\frac12,\frac12)$ and transforms under supersymmetry as
\begin{equation}
\delta P_L {\cal G} = \overline{\phi ^0} P_L \epsilon + \cdots \, ,
\end{equation}
where dots refer to terms with fermionic component fields. We obtain therefore the expressions
\begin{equation}
\begin{aligned}
   (\mathcal{R})_{v}^a= & \frac14{\rm i}\overline{{\cal  G}}\gamma _*\gamma^a  {\cal  G}+\dots\,,\\
 (\mathcal{R})_{P_L\lambda} =&{\rm i}\phi^0 P_L{\cal G} +\dots\,,\\
(\mathcal{R})_\tD =&2\phi^0\overline\phi^0+\frac12\overline{{\cal G}}\slashed{\cal D}{\cal G}+\left[
 \frac{1}{ 2\phi^0} \overline{\cal G} P_L {\cal G} F^0-\sqrt{2}\,\overline \Omega^0P_L{\cal G}+\hc\right]+\dots\,.
\label{finalcomponents}
\end{aligned}
\end{equation}
The leading order contributions in the fermions to the second term in (\ref{newDterm}) are obtained by inserting \eqref{finalcomponents} into \eqref{prodVL}:
\begin{align}
\label{XIL}
e^{-1} \left[\mathcal{R}\, {\tD}\right]_D =\phi^0 \overline{\phi^0}\, \tD
&+\frac{1}{4}\tD\left[\overline{\mathcal{G}}  \slashed{\mathcal{D}} \mathcal{G} + \overline{\cal G} \left(P_L \frac{ F^0}{\phi^0}+P_R\frac{\overline{ F^0}}{\overline{\phi^0}}\right){\cal G}- 2\sqrt 2 \overline {\mathcal G} \Omega^0
+ \overline \psi\cdot\gamma \left(P_L\phi ^0+P_R\overline{\phi ^0}\right) \mathcal{G}\right]\nonumber\\
&+\frac{\rmi}{8} \overline{\mathcal{G}} \gamma _* \gamma^a\mathcal{G}\,  \mathcal{D}^bF_{ab}
- \frac{\rmi}{2} \overline{\mathcal{G}}\left( P_L\phi^0-P_R\overline{\phi ^0}\right) \slashed{\mathcal{D}}\lambda
+  \text{4-fermions} \, .
\end{align}
Note that the terms without the gauge curvature in the Poincar\'{e} gauge $\phi ^0=\kappa ^{-1}$, $\Omega ^0=0$, where ${\cal G} = -2\rmi \frac{1 }{\kappa\tD}\gamma _*\lambda+\ldots $, simplify to
\begin{align}
- \kappa ^2\xi \left. \left[\mathcal{R}\, {\tD}\right]_D \right|_\poinc={\cal L}_{\rm FI}+{\cal L}_{\rm mix}+
 +\kappa e \frac{\xi }{\tD} \overline{\lambda } \left(P_L  F^0+P_R\overline{ F^0}\right)\lambda+  \text{4-fermions + gauge field terms} \, ,
\label{RDsimplified}
\end{align}
where ${\cal L}_{\rm FI}$ and ${\cal L}_{\rm mix}$ are the expressions  (\ref{LFI}) and  (\ref{LMIX}), which appeared in the Freedman model. Note, however, that before gauge fixing the first term is
\begin{equation}
 {\cal L}_{\rm FI, new}= \kappa ^2\phi ^0 \overline{\phi ^0}{\cal L}_{\rm FI}\,.
 \label{FInew}
\end{equation}
This prefactor, which is here just 1, is going to be relevant when discussing matter couplings in Section~\ref{sec:matt_coup}.

At this point it is instructive to understand the deeper origin of the difference between the model presented here and the model of Freedman discussed in the previous section.
The term in (\ref{XIL}) linear in $F_{\mu \nu }$ and in the gravitino, namely the first correction to (\ref{RDsimplified}), is
\begin{equation}
\label{Nterm3}
{\cal L}_{\rm diff}= - e \xi  \frac{1 }{2\tD}\, \overline \psi_\mu \gamma^{\mu\nu \rho} \lambda \,\, F_{\nu\rho}  \,.
\end{equation}
It is now the variation $\delta \lambda = \frac{1}{2}\rmi\gamma _*\epsilon \tD$ that cancels (\ref{Ncrucial}) and thus replaces the contribution of the modified gravitino covariant derivatives.
We observe that such term, after the $\tD$ field equation are used, does not vanish in the $\xi \rightarrow 0$ limit:
\begin{equation}
\label{Nterm3xi}
{\cal L}_{\rm diff}= - e   \frac{1 }{2}\, \overline \psi_\mu \gamma^{\mu\nu \rho} \lambda \,\, F_{\nu\rho}  \, ,\qquad \delta \lambda = \ldots + \frac{\rmi}{2} \xi  \gamma_* \epsilon\,.
\end{equation}
The complete Lagrangian \eqref{newDterm} still is ill defined in this limit.
\bigskip

We proceed now to couple the Lagrangian \eqref{newDterm} to the pure supergravity sector.
In the superconformal setup the minimal model is described by
\begin{equation}
\label{LDtermnew}
{\cal L}
 = -3 \left[ \phi^0 \overline{\phi^0}\right]_{D}
+  \left[(\phi^0)^3 \kappa m_{3/2} \right]_{F}
+ {\cal L}_\text{NEW} \, ,
\end{equation}
where $m_{3/2}$ is a real constant which is going to be identified with the gravitino mass.
Notice that, since there is no R-symmetry gauging, terms are allowed that explicitly break the R-symmetry. This is the reason why a non-vanishing gravitino mass can be consistently included within our model.
Once we gauge fix and integrate out all the auxiliary fields, one can notice that supersymmetry is spontaneously broken by $\langle \text{D} \rangle = \xi$ and the goldstino is identified with the gaugino $\lambda$.
Since the vacuum expectation value of $\text{D}$ is not allowed to vanish, supersymmetry cannot be restored and a smooth $\xi \rightarrow 0$ limit does not exist, once the Lagrangian is written on-shell.

The complete component form of the off-shell Lagrangian, after gauge fixing the conformal symmetry setting $\phi^0=\kappa^{-1}$ and
in the gauge in which the goldstino is gauge-fixed to vanish, namely the unitary gauge, reads
\begin{equation}
\begin{aligned}
\label{LDnewOFF}
e^{-1}\mathcal{L}
=& \frac1{2\kappa ^2}\left(R(\omega(e,\psi))-\overline{\psi_\mu} \gamma^{\mu\nu\rho}\left(\partial_\nu+\frac14 {\omega_\nu}^{ab}(e,\psi)\gamma_{ab}\right)\psi_\rho+6 A_a A^a
+ m_{3/2} \overline \psi_\mu\gamma^{\mu\nu}\psi_\nu\right)\\
&-\frac14 F_{\mu\nu}F^{\mu\nu}
-3 F^0 \overline{F^0}
+\frac{6}{\kappa }m_{3/2}\text{Re}(F^0)+\frac12 \text{D}^2-\xi\text{D} \, .
\end{aligned}
\end{equation}
The equations of motion for the auxiliary fields give in the unitary gauge: $F^0=\overline{F^0}=\kappa ^{-1}m_{3/2}$, $\text{D}=\xi$ and $A_\mu=0$.
The on-shell Lagrangian is therefore
\begin{equation}
\label{LLcomp}
\begin{aligned}
e^{-1}\mathcal{L}
=& \frac1{2\kappa ^2}\left(R(\omega(e,\psi))-\overline{\psi_\mu} \gamma^{\mu\nu\rho}\left(\partial_\nu+\frac14 {\omega_\nu}^{ab}(e,\psi)\gamma_{ab}\right)\psi_\rho+ m_{3/2} \overline \psi_\mu\gamma^{\mu\nu}\psi_\nu\right)\\
&-\frac14 F_{\mu\nu}F^{\mu\nu}
-\left(\frac12 \xi^2-\frac{3}{\kappa ^2}(m_{3/2})^2\right).
\end{aligned}
\end{equation}

A notable difference between the model \eqref{LLcomp} and the standard D-term model of Freedman \eqref{FRonshell}
is that the vector $v_\mu$ is not appearing in the covariant derivative of the gravitino in \eqref{LLcomp} as connection for the R-symmetry, since the latter is not gauged.
In addition notice that, even though we started from a Lagrangian with a complicated non-linear term,
there are no non-linearities in the bosonic sector
and neither in the complete action when it is written in the unitary gauge.
In this model, finally, the supersymmetry breaking scale is proportional to $\sqrt \xi$ and the unitarity bound for the gravitino
mass derived in \cite{Deser:2001us} is always respected.

\section{Matter couplings and uplift}
\label{sec:matt_coup}

We are going now to couple the Lagrangian \eqref{newDterm} to standard supergravity together with a set of chiral multiplets $\phi^I=\{\phi^0,\phi^i\}$.
The multiplet $\phi^0$ still serves as conformal compensator with Weyl and chiral weights $(1,1)$,
while the matter multiplets $\phi^i$ have vanishing weights.
We therefore consider the couplings
\begin{equation}
\label{LDnewSUGRA}
{\cal L} =  -3 \left[ \phi^0 \overline{\phi^0} \text{e}^{-K(\phi^i,\overline{\phi^i})/3} \right]_{D}
+  \left[(\phi^0)^3 W(\phi^i) \right]_{F}
+ {\cal L}_\text{NEW} \, ,
\end{equation}
where $K(\phi^i,\overline{\phi^i})$ is the K\"ahler potential and $W(\phi^i)$ is the superpotential of the chiral model.
The properties of the Lagrangian \eqref{LDnewSUGRA}, without the new term ${\cal L}_\text{NEW}$, can be found in \cite{Freedman:2012zz}.\footnote{In comparison to that reference, our $K$ is $\kappa ^2{\cal K}$, our $W$ is $\kappa ^3 W$, $\phi ^0= \kappa ^{-1}y$, and the $\phi ^i$ are the $z^\alpha $.}
In the next section we are going to recast the model \eqref{LDnewSUGRA} into an equivalent one written in terms of constrained multiplets.

We concentrate now on the bosonic sector of \eqref{LDnewSUGRA}.
As anticipated,
the contribution of ${\cal L}_\text{NEW}$ in the chiral model \eqref{LDnewSUGRA} generates a positive definite term in the scalar potential.
After gauge fixing of superconformal symmetry, setting $\phi^0=\kappa^{-1} \rme^{K/6}$, and once the auxiliary fields are integrated out,
the pure bosonic sector of the theory reads
\begin{equation}
\label{bos1}
{e}^{-1} {\cal L}^{(B)} = \frac1{2\kappa ^2} R
- \frac14 F_{\mu \nu} F^{\mu \nu}
- \frac{1}{\kappa ^2}g_{i \jb}\, \partial \phi^i \partial \overline \phi^{\jb}
- {\cal V}  \, ,
\end{equation}
with the scalar potential taking the form
\begin{equation}
\label{scalarP}
{\cal V} = \kappa ^{-4}{\rme}^{K} \left( |\nabla _i W|^2 - 3 |W|^2 \right)
+ \frac{\xi^2}{2} \text{e}^{2K/3} \, ,
\end{equation}
where $\nabla _i W = \partial_i W + K_i W$ and $|\nabla _i W|^2= g^{i\jb} \nabla _i W \overline{\nabla }_{\jb} \overline W$.
Note that the factor in the $\xi $ term originates from the prefactor in (\ref{FInew}).
We remind the reader that, within this setup, supersymmetry always has to be spontaneously broken, albeit linearly-realized, and the auxiliary field of the abelian vector multiplet has to be non-vanishing on the vacuum, otherwise the model becomes ill defined.

As a simple application, let us couple the new term to a single chiral multiplet $T$
and investigate the resulting theory.
In particular we choose the K\"ahler potential and superpotential to be of the type
studied in \cite{Ferrara:2014kva,Kallosh:2014wsa,Bergshoeff:2015jxa},
but we make no further assumptions and we are not including any constrained multiplet.
We have therefore
\begin{equation}
\label{TKW}
K(T, \overline T) = -3 \, \text{log}\left( T + \overline T \right) , \quad W(T) = W_0 + A \, \text{e}^{-a T} \, ,
\end{equation}
where $A$ and $a$ are constants.
The full bosonic sector is given by \eqref{bos1}
but, due to the new contribution proportional to $\xi$, the scalar potential \eqref{scalarP} has the form
\begin{equation}
{\cal V} =  \kappa ^{-4}\text{e}^{K} \left( |\nabla _i W|^2 - 3 |W|^2 \right) + \frac{\xi^2}{2 \left( T + \overline T \right)^2} \, .
\end{equation}
This specific setup has been studied in \cite{Ferrara:2014kva,Kallosh:2014wsa,Bergshoeff:2015jxa} with the use of constrained superfields. It describes the impact of a $\overline{D3}$ probe brane on the scalar potential of standard supergravity.

Interestingly, the model \eqref{bos1} not only has such a
very specific form for the scalar potential, but also contains an abelian gauge vector. It matches therefore exactly the field content of the effective theory for the $\overline{D3}$ brane at strong warping,
where $\xi$ is proportional to the warping factor \cite{Kachru:2003sx,Bandos:2016xyu}.

% section new_d_term (end)

\section{Emergence of non-linear realizations}
\label{sec:d_term_nilpotent}

In this section we make explicit use of non-linear realizations of supersymmetry in order to recast the model \eqref{LDnewSUGRA} into an equivalent one, in which the relation of our work to the formalism of constrained multiplets is manifest.
In particular, the chiral goldstino multiplet $X$ is emerging and we are going to show how a K\"ahler potential and a linear superpotential in $X$ are generated. In other words, the D-term breaking model, where supersymmetry is linearly realized, is recast into an F-term breaking one, where supersymmetry is non-linearly realized.

The key ingredient for the entire construction is a chiral multiplet $X=\{X,\Omega^X,F^X\}$, with Weyl and chiral weights $(1,1)$ and that is constrained to be nilpotent \cite{Rocek:1978nb,Lindstrom:1979kq,Casalbuoni:1988xh}
\begin{equation}
\label{X2}
X^2=0 \,   \qquad \Longleftrightarrow \qquad X = \frac{\overline \Omega^X P_L \Omega^X }{2 F^X} \,.
\end{equation}
In particular, for this constraint to be imposed consistently, the auxiliary field $F^X$ has to be non-vanishing on the vacuum.
For the rest of the discussion, when referring to the multiplet $X$, we always assume it satisfies the constraint \eqref{X2}.
Once the scalar $X$ is replaced by the composite expression $\frac{\overline \Omega^X P_L\Omega^X}{2 F^X}$, local supersymmetry becomes non-linearly realized. In general, when other multiplets are present in the theory, one can impose constraints on them in order to eliminate specific component fields \cite{Ferrara:2016een,DallAgata:2016syy}.

As a first step in our analysis we organize the degrees of freedom of the vector multiplet in two parts,
one of which is going to contain the vector component field, while the other is going to contain the goldstino and the auxiliary field, which breaks supersymmetry.
As shown in \cite{Cribiori:2017ngp}, this can be accomplished by parametrizing unconstrained multiplets in terms of constrained ones.
To this purpose, beside the nilpotent chiral multiplet $X$, we introduce another real vector multiplet $\tilde V = \{\tilde V,\,\tilde \zeta,\,\tilde{\mathcal{H}},\,\tilde v_\mu,\,\tilde \lambda,\, \tilde{\text{D}}\}$
that satisfies the constraints\footnote{Similar as in superspace, we can define the multiplets that have $P_L\tilde \lambda$ and $\tilde \tD$ as lowest components by respectively $\tilde \lambda _\alpha=\text{T}({\cal D}_\alpha \tilde V)$ and ${\cal D}^\alpha  \tilde \lambda _\alpha$. These operations are consistent with the Weyl and chiral weights as summarized in appendix B of \cite{Ferrara:2016een}, following \cite{Kugo:1983mv}.}
\begin{equation}
\label{DLconstr}
X \overline X\, \tilde{\text D} = 0  \ , \qquad X\, P_L\tilde{\lambda}=0 \, .
\end{equation}
The first constraint eliminates the highest component $\tilde{\text{D}}$ while the second eliminates the gaugino $\tilde \lambda$. These component fields are both expressed in terms of the other degrees of freedom in the theory.
For a recent discussion in global supersymmetry in a setup where the vector is eliminated see \cite{Benakli:2017yar}.

Notice that, when we write  (\ref{X2}) as\footnote{Such a splitting of $X$, but then in terms of two chiral multiplets $X={\cal A}_+Z$, is discussed in the appendix A. See $e.g.$ the first of (\ref{ZA+-}) with (\ref{ZAbosonic}). In \cite{Ivanov:1982bpa} another splitting is presented where the
linear multiplet is split in multiplets transforming under the
standard non-linear realizations of supersymmetry. In \cite{Komargodski:2009pc,Bandos:2016xyu} for example other similar splittings are presented.
However these are not the splittings we use here.}
\begin{equation}
  X = \frac{1}{2}F^X \overline{\chi }P_L\chi \,,\qquad  P_L\chi \equiv \frac{P_L\Omega ^X}{F^X}\,,
 \label{defchi}
\end{equation}
the supersymmetry transformation of  $P_L\chi $ is only a function of $P_L\chi $ and not of $F^X$.
 Therefore $F^X$ is an overall nonzero factor in (\ref{DLconstr}) and the components $P_L\tilde{\lambda}$ and $\tilde{\text D}$ depend on the components of $X$ only through $P_L\chi $ (and $P_R\chi $ for $\tilde{\text D}$).

We are now in the position to parametrize the vector multiplet $V$ in terms of the constrained multiplet $\tilde V$ and $X$
\begin{equation}
\label{VtV}
V = \tilde V
+ \frac{1}{2\sqrt 2}\frac{X \overline X}{\phi ^0\overline{\phi ^0}} \left(\frac{1}{t}+\frac{1}{\overline{t}}\right)\,,\qquad t\equiv \frac{\text{T} \left( \overline X  \right)}{(\phi ^0)^2}\,.
\end{equation}
This parametrization is the generalization within supergravity of the one presented in \cite{Cribiori:2017ngp}.
Note that in this expression $F^X$ appears only in the combination $\chi $ and it appears in the second term, due to the proportionality of $X$ and $\overline{\text T}(X)$ with $F^X$, in the form
\begin{equation}
  f\equiv \Re \left(F^X\frac{\phi ^0}{\overline{\phi ^0}}\right)\,.
 \label{deff}
\end{equation}
The expression in the bracket indeed has conformal weights $(2,0)$, so that one can take the real part. In other words, $f$ is the real part of the lowest component of $\phi ^0\, \overline{\phi ^0}\,t$.
Notice also that the bosonic part of the $D$-component of $V$ in (\ref{VtV}) can be identified as
\begin{equation}
  \tD = \sqrt{2}f+\mbox{fermionic terms}\,.
 \label{tDisf}
\end{equation}
To sum up, as a consequence of \eqref{VtV}, the vector $v_\mu$ is replaced by $\tilde v_\mu$ plus some function of $\chi $ and $f$, while $\lambda$ and $\text D$ are given entirely in terms of $\chi $ and $f$.

In the following, instead of the known Wess--Zumino gauge choice, it is going to be convenient to adopt the modified gauge condition proposed in \cite{Komargodski:2009rz}, namely
\begin{equation}
\label{XV0}
X \tilde V=XV = 0 \, ,
\end{equation}
which means that the components $\tilde V,\tilde \zeta$ and $\tilde{\mathcal{H}}$ of $\tilde V$ are removed from the spectrum and expressed as functions of the goldstino and of the remaining degrees of freedom.
To preserve this gauge choice, the chiral multiplet ${\cal B}$ entering the gauge transformations \eqref{Btransform} is required to satisfy $X {\cal B} = X \overline{\cal B}$, which implies that the only independent field in $\mathcal{B}$ is a real scalar in the lowest component.

In other words after these redefinitions, the independent $4+4$ field components, apart from the pure supergravity fields and the compensator $\phi ^0$, can be embedded into
\begin{equation}
  \left\{\tilde v_\mu,\,\chi ,\,f\right\}\,,
 \label{indepfields}
\end{equation}
where $\tilde v_\mu$ is a gauge vector and $f$ is real.

Due to the properties  (\ref{X2}),  (\ref{DLconstr}) one can then rewrite the chiral multiplet $\overline\lambda P_L\lambda $ as\footnote{The main tools in these calculations are the identities $\text{T}(Z\,Y)= Z\, \text{T}(Y)$ for $Z$ a  chiral multiplet and any $Y$, and that the nilpotent properties imply $\overline{X}\text{T}\left(\overline{X}\,Y\right)=\overline{X}\text{T}\left(\overline{X}\right)\,Y$.\label{fn:tools}}
\begin{equation}
  \overline\lambda P_L\lambda= \overline{\tilde \lambda} P_L \tilde \lambda -\frac{1}{2}X\, T\left[\overline X\left(\frac{(t+\overline{t})^2}{t\,\overline{t}}
    \right)\right]\,.
 \label{rewritelamlam}
\end{equation}
The $F$-component of this multiplet has pure bosonic part:
\begin{equation}
  \left(\overline\lambda P_L\lambda\right)_F=\tilde {F}^-\cdot \tilde {F}^--2f^2+\mbox{fermionic terms}\,,\qquad \tilde {F}_{\mu \nu }= 2\partial _{[\mu }\tilde v_{\nu ]}\,,
 \label{FcomponenttildeW}
\end{equation}
which in fact is similar to (\ref{lambda2Fcomp}), after using (\ref{tDisf}). Since for any real ${\cal C}$ of Weyl weight~2:
\begin{equation}
  [{\cal C}]_D= \frac12[\text{T}({\cal C})]_F\,,
 \label{DactionfromF}
\end{equation}
 (\ref{rewritelamlam}) implies for the action
\begin{equation}
\begin{aligned}
\label{W2calc}
-\frac14 [\overline\lambda P_L\lambda]_F = -\frac14 [\overline{\tilde \lambda} P_L \tilde \lambda ]_F
+\frac14\,
\left[X\overline X\left(\frac{(t+\overline{t})^2}{t\,\overline{t}}
    \right)
    \right]_D \, .
\end{aligned}
\end{equation}
 Note that  (\ref{DLconstr}) implies that $P_L\tilde \lambda $ is proportional to $P_L\Omega ^X$ and therefore both terms in (\ref{rewritelamlam}) are proportional to $X$. Thus $w^2\overline{w}^2$ is proportional to $X\overline X$ and with the methods of footnote \ref{fn:tools} one can then obtain
\begin{equation}
  \frac{w^2 \overline{w}^2}{\text{T}(\overline{w}^2)\overline{\text T}(w^2)}=\frac{X\,\overline X}{\text{T}(\overline{X})\overline{\text T}(X)}\,.
 \label{woverTisXoverT}
\end{equation}
The left-hand side appears in ${\cal R}$ in (\ref{Rmultiplet}). Concentrating now on the action term $\left[\mathcal{R}\, {\tD}\right]_D$, the proportionality of $\mathcal{R}$ with $X\,\overline X$ discussed above implies that only some parts of $\tD$ have to be taken into account. The $D$-component of
 the first term in (\ref{VtV}) can be omitted and for the $D$-component of the second term the $X\overline X$ is replaced by $2F^X\overline F^X= 2\overline{\text T}(X)\text{T}(\overline{X})$. Thus only the explicit term in (\ref{tDisf}) survives:
\begin{equation}
  X\overline{X}\,\tD = \frac{1}{\sqrt{2}}X\overline{X}\frac{\text{T}(\overline{X})\overline{\text T}(X)}{\phi ^0\overline{\phi ^0}}\left(\frac{1}{t}+\frac{1}{\overline{t}}\right)\,,
 \label{XbarXD}
\end{equation}
and (\ref{woverTisXoverT}) and  (\ref{XbarXD}) imply thus
\begin{equation}
  \mathcal{R}\, {\tD} = \frac{1}{\sqrt{2}}X\,\overline X\left(\frac{1}{t}+\frac{1}{\overline{t}}\right)\,.
 \label{newterminX}
\end{equation}
The theorem \cite{Cecotti:1987sa}, written in our notations in (7.7) of \cite{Ferrara:2016een}, then implies
\begin{equation}
  \left[\mathcal{R}\, {\tD}\right]_D =  \frac{1}{\sqrt{2}}\left[X \frac{\text T (\overline X)}{t}\right]_F = \frac{1}{\sqrt{2}}\left[(\phi ^0)^2\,X \right]_F \,.
 \label{WX}
\end{equation}
A linear superpotential term for $X$ has therefore been generated.

Combining everything, the Lagrangian (\ref{newDterm}) is
\begin{equation}
  {\cal L}_{\rm NEW}= -\frac14 \left[\overline {\tilde \lambda} P_L\tilde \lambda\right]_F
  +\frac14\,\left[X\overline X\left(\frac{(t+\overline{t})^2}{t\,\overline{t}}\right)\right]_D
  - \frac{1}{\sqrt{2}}\kappa ^2\xi \left[(\phi ^0)^2\,X \right]_F \,.
 \label{LnewX}
\end{equation}

The second term has an on-shell equivalence with the action $[X\overline X ]_D$. Similar to what has been done in global supersymmetry in \cite{Cribiori:2017ngp} one can add to this Lagrangian a contribution
\begin{equation}
  {\cal L}_{\rm add}=-\frac14\,\left[X\overline X\left(\frac{(t-\overline{t})^2}{t\,\overline{t}}\right)\right]_D\,.
 \label{Ladd}
\end{equation}
This term is proportional to $t-\overline t$, which does not appear anywhere in (\ref{LnewX}), and thus it is on-shell trivial. A more detailed argument is given in the appendix A, where the degrees of freedom of $\chi $ and $f$ (or $\Re t$) are encoded in the multiplets $Z$ and ${\cal A}_1$, while the degree of freedom that appears only in (\ref{Ladd}) is contained in ${\cal A}_2$.

Adding (\ref{Ladd}) to  (\ref{LnewX}) we thus obtain the simple result:
\begin{equation}
\label{KX}
{\cal L}_{\rm NEW}+{\cal L}_{\rm add}=
-\frac14 \left[\overline{\tilde \lambda}P_L\tilde\lambda\right]_F+[X\overline X ]_D - \frac{1}{\sqrt{2}}\kappa ^2\xi \left[(\phi ^0)^2\,X \right]_F\,.
\end{equation}
A canonical K\"{a}hler potential has also been generated for the nilpotent multiplet $X$.

When the supergravity action and matter couplings are included, the original Lagrangian \eqref{LDnewSUGRA} gets the final form
\begin{equation}
\label{LDtermnewX}
{\cal L} = -3 \left[ \phi^0 \overline{\phi^0} \text{e}^{-K(\phi^i,\overline{\phi^i})/3} \right]_{D}
+ \left[X\overline X\right]_D
+  \left[- \frac{1}{\sqrt{2}}\kappa ^2\xi (\phi^0)^2 X  + (\phi^0)^3 W(\phi^i) \right]_{F}
-\frac14  \left[\overline{\tilde \lambda} P_L \tilde \lambda \right]_{F}  \, .
\end{equation}
This model describes the interactions of supergravity with the real vector $\tilde v_\mu$, the fermion $\Omega^X$
and the scalars and fermions of the matter chiral multiplets $(\phi^i,\Omega^i)$. The breaking is manifestly F-term. We stress the fact that
supersymmetry has to be broken and there always has to be a contribution to the breaking from the auxiliary field of $X$,
otherwise the model is ill defined.
The multiplets appearing in \eqref{LDtermnewX} satisfy the constraints
\begin{equation}
\label{constrs}
X^2=0 \,, \qquad X \overline X\, \tilde{\text D}= 0  \,, \qquad X\, P_L\tilde{\lambda}=0 \, ,\qquad \tilde{\text D}={\cal D}^\alpha \tilde{\lambda}_\alpha \,,
\end{equation}
while the matter chiral multiplets $\phi^i$ are unconstrained.
Despite its simple form, the model \eqref{LDtermnewX} is equivalent to \eqref{LDnewSUGRA} and its pure $X$ sector coincides with the one of \cite{Ferrara:2014kva,Kallosh:2014wsa,Bergshoeff:2015jxa,Antoniadis:2014oya,DallAgata:2014qsj,Dudas:2015eha,Bergshoeff:2015tra,Hasegawa:2015bza,Kuzenko:2015yxa,Bandos:2015xnf}.

We can recast \eqref{LDtermnewX} into a more familiar form, defining the chiral multiplet $S= X/\phi ^0$ with vanishing weights. We obtain then the Lagrangian
\begin{equation}
\label{LDnewSUGRAconstr}
{\cal L} = -3 \left[ \phi^0 \overline{\phi^0} \text{e}^{-\hat K/3} \right]_{D}
+  \left[(\phi^0)^3 \, \hat W \right]_{F}
-\frac14 \left[ \overline{\tilde \lambda} P_L \tilde \lambda \right]_F  \, ,
\end{equation}
where the K\"ahler potential is
\begin{equation}
\hat K =  -3 \, \text{log}\left( \text{e}^{-K(\phi^i,\overline{\phi^i})/3} - \frac13 S \overline S \right)
=  \, K(\phi^i,\overline{\phi^i}) + S \overline S \, \text{e}^{K(\phi^i,\overline{\phi^i})/3}  \, ,
\end{equation}
and the superpotential is
\begin{equation}
\hat W = W(\phi^i) - \kappa ^2\xi S / \sqrt 2 \, .
\end{equation}
As a simple application, let us couple the new term to a single chiral superfield $T$
and investigate the resulting theory.
In particular, for the $T$ sector we choose the K\"ahler potential and superpotential given in \eqref{TKW}.
Due to the coupling to our new term, the model takes the form
\begin{equation}
\hat K = -3 \, \text{log}\left( T + \overline T - \frac13 S \overline S \right) , \quad \hat W = W_0 + A \, \text{e}^{-a T} -\frac{1}{\sqrt{2}} \kappa ^2\xi S \, .
\end{equation}
The relation between our model and the ones in \cite{Ferrara:2014kva,Kallosh:2014wsa,Bergshoeff:2015jxa} describing an $\overline{D3}$ uplift is now manifest.
We have to stress however that we started from the Lagrangian \eqref{LDnewSUGRA}
and therefore we have not introduced any constrained multiplet in the theory beforehand.

\section{Conclusions and discussion}
\label{ss:concl}

We showed how an action with a gauge multiplet can be constructed that contains a Fayet--Iliopoulos term without R-symmetry gauging. This action has no non-linearities in the bosonic sector or in unitary gauge, but is highly non-linear in the fermions. For consistency the FI constant $\xi $ should be non-vanishing and the model thus describes broken supersymmetry. It can be included in a full model with matter multiplets, a K\"{a}hler manifold and a superpotential,
even though the $\xi$ term explicitly breaks K\"{a}hler invariance.\footnote{This is at least so for the formulation with constant $\xi $. As mentioned below, $\xi $ could be taken field-dependent, which can restore a modified K\"{a}hler transformation in which $\xi $ transforms under K\"{a}hler transformations.}
In particular, the impact of a $\overline{D3}$ probe brane on the scalar potential of standard supergravity is nicely described in this setup. We took in this paper a constant parameter $\xi $ and a unit gauge kinetic function for the action of the gauge multiplets. Note, however, that once matter multiplets are introduced, the parameter $\xi$ in \eqref{newDterm} can become a real field-dependent function $\xi(\phi^i,\overline{\phi^i})$. The gauge kinetic function, which is set to unit in \eqref{newDterm},
can also be a holomorphic function of the chiral multiplets.

In the last few years it was found that models with supersymmetry breaking have a nice description in terms of constrained multiplets \cite{Rocek:1978nb,Ivanov:1978mx,Lindstrom:1979kq,Samuel:1982uh,Casalbuoni:1988xh,Komargodski:2009rz,Kuzenko:2017zla}. A systematic description for rewriting regular superfields in terms of such constrained building blocks for broken global supersymmetry has been developed in \cite{Cribiori:2017ngp}. We upgraded ingredients of that approach to the local superconformal tensor calculus to rewrite the new FI model. We found that after suitable redefinitions, our model can be written in terms of a constrained gauge multiplet $\tilde \lambda $ (with gauge field $\tilde v_\mu $) and the nilpotent chiral multiplet $X$ with a superpotential, describing the supersymmetry breaking.

As a final comment, we observe that the model we are presenting enjoys a duality invariance of the type discussed in \cite{Tseytlin:1996it,Bagger:1996wp}, which implies the electric-magnetic duality for the $U(1)$ gauge field.
This is an on-shell duality and, in order to make it manifest, one can consider the Lagrangian \eqref{LDtermnewX} and notice that the second constraint in \eqref{constrs} can be relaxed on-shell as, after reintroducing an independent $\rm{\tilde{D}}$, the solution of its equations of motion is satisfying also the aforementioned constraint. The third constraint in \eqref{constrs} can then be implemented with a Lagrange multiplier and, following the procedure of \cite{Bagger:1996wp}, one can obtain the Lagrangian \eqref{LDtermnewX} where the multiplet $P_L\tilde \lambda$ is replaced with its dual.
The interested reader can find more details in appendix B.
This procedure shows that the Lagrangian \eqref{LDnewSUGRA} enjoys an electric-magnetic duality.

% section d_term_nilpotent (end)

\bigskip
\section*{Acknowledgments}

\noindent We would like to thank Gianguido Dall'Agata, Christoph Roupec and Timm Wrase for discussions. N.C. thanks the KU Leuven for hospitality. This work is supported
in part by the Interuniversity Attraction Poles Programme
initiated by the Belgian Science Policy (P7/37), and in part by
support from the KU Leuven C1 grant ZKD1118 C16/16/005.
N.C. is supported in parts by the Padova University Project CPDA144437. The work of M.T. is supported by the FWO odysseus grant
G.0.E52.14N. M.T. would like to thank the University of Amsterdam for its hospitality during which part of this work has been done. Furthermore M.T. is grateful to the Delta ITP for financing his stay.

\appendix
\section{Parametrization of the \texorpdfstring{$X$}{X} superconformal multiplet}
This appendix contains the local superconformal version of the equations developed in rigid superspace in \cite{Komargodski:2009rz,Cribiori:2017ngp}.
We start from the nilpotent chiral multiplet $X$ with Weyl weight~1. From this one we define
\begin{equation}
  Z= \frac{X}{\tT(\overline{X})}\tT\left(\frac{(\overline{\phi ^0})^2\overline{X}}{\overline{\tT}(X)}\right)\,,
 \label{defZ}
\end{equation}
which is also a chiral multiplet with Weyl weight~1. Since it is proportional to $X$, this multiplet is nilpotent and its lowest component is eliminated in terms of the others.  Calculating $Z\tT(\overline{Z})$ using the methods of footnote \ref{fn:tools} we obtain
\begin{equation}
  Z\,\tT(\overline{Z})= ({\phi ^0})^2\, Z\,,
 \label{ZTbZ}
\end{equation}
which implies that also the auxiliary field component of $Z$ is defined in function of the fermionic component. Therefore $Z$ contains only one fermion as independent degree of freedom.

Moreover $X$ and $Z$ satisfy
\begin{equation}
  X\, \frac{\overline{\tT}(Z)}{\overline{\tT}(X)}= Z\,.
 \label{identityZchiralantichiral}
\end{equation}
Defining, the following chiral superfields of chiral weight~0:
\begin{align}
 & {\cal A}_1= {\cal A}_+ + {\cal A}_-\,,\qquad {\cal A}_2 = -\rmi\left({\cal A}_+ - {\cal A}_-\right)\,,\nonumber\\
 &{\cal A}_+=\frac{1}{\tT(\overline{Z})}\tT\left(\overline{Z}\frac{\overline{\tT}(X)}{\overline{\tT}(Z)}\right)\,,\qquad
 {\cal A}_-=\frac{1}{\tT(\overline{Z})}\tT\left(\overline{Z}\frac{\tT(\overline{X})}{\tT(\overline{Z})}\right)= \frac{\tT(\overline{X})}{\tT(\overline{Z})}\,,
 \label{defcalA}
\end{align}
we find using (\ref{identityZchiralantichiral})
\begin{equation}
  Z{\cal A}_+=Z\overline{{\cal A}}_-=X\,,\qquad Z{\cal A}_-=Z\overline{{\cal A}}_+=Z\frac{\tT(\overline{X})}{\tT(\overline{Z})}\,,
 \label{ZA+-}
\end{equation}
such that
\begin{equation}
  Z\left({\cal A}_1-\overline{{\cal A}}_1\right)=0\,,\qquad Z\left({\cal A}_2-\overline{{\cal A}}_2\right)=0\,.
 \label{ZA1A2}
\end{equation}
This implies that ${\cal A}_1$ and ${\cal A}_2$ contain only one real scalar in the lowest component as independent
degree of freedom. We have therefore decomposed the nilpotent goldstino multiplet $X$ into three
constrained multiplets, one pure goldstino multiplet $Z$ and two constrained chiral multiplets ${\cal A}_1$ and~${\cal A}_2$.

To compare these with the objects in section \ref{sec:d_term_nilpotent}, we extract from these objects the leading terms:
\begin{align}
  Z & = \frac{X}{t}+ \mbox{ 4-fermion terms}\,,\nonumber\\
     {\cal A}_+  & = \overline{t} +\mbox{ fermionic terms}\,,\qquad {\cal A}_-   = t +\mbox{ fermionic terms}\,,
\label{ZAbosonic}
\end{align}
Therefore ${\cal A}_1$ contains the real part of the auxiliary field that breaks supersymmetry, and ${\cal A}_2$ contains its imaginary part.

Since $t$ and $\overline{t}$ appear in (\ref{LnewX}) and  (\ref{Ladd}) multiplied by $X\overline X$, being 4-fermion terms, we have also
\begin{equation}
  X\overline X\left(\frac{(t+\overline{t})^2}{t\,\overline{t}}\right) = Z\overline{Z}{\cal A}_1^2 \,,\qquad X\overline X\left(\frac{(t-\overline{t})^2}{t\,\overline{t}}\right) =- Z\overline{Z}{\cal A}_2^2\,.
 \label{rewriteXtinA}
\end{equation}
In fact, using the decomposition in (\ref{ZA+-}), we also have directly
\begin{equation}
  X\overline{X}= Z\overline{Z}{\cal A}_+{\cal A}_- = Z\overline{Z}\left({\cal A}_1^2 + {\cal A}_2^2\right)\,.
 \label{XbarXinA1A2}
\end{equation}
Similarly, by the definitions in (\ref{defcalA}) and using  (\ref{ZTbZ}) and (\ref{identityZchiralantichiral}), we have
\begin{equation}
 (\phi _0)^2 X= (\phi ^0)^2 Z{\cal A}_+=\tT\left(\overline{Z}X\right)\,,\qquad (\phi ^0)^2 Z{\cal A}_- =\tT\left(Z\overline{X}\right)\,.
 \label{phi0XinA}
\end{equation}
Due to the property \cite{Ferrara:2016een} $\left[\tT({\cal C})\right]_F = \left[\tT(\overline{{\cal C}})\right]_F$, the $F$-term density of both expressions is equal and thus
\begin{equation}
  \left[ (\phi _0)^2 X\right]_F=\frac{1}{2} \left[ (\phi ^0)^2 Z{\cal A}_1\right]_F\,.
 \label{phi0XA1}
\end{equation}
Since ${\cal A}_2$ depends on the imaginary part of $t$, which does not appear in ${\cal A}_1$ (due to the constraints (\ref{ZA1A2})), the last term in (\ref{XbarXinA1A2}), which is equal to ${\cal L}_{\rm add}$ (\ref{Ladd}) as a consequence of (\ref{rewriteXtinA}), is on-shell trivial.

\section{The electric-magnetic duality of the new D-term}

In this appendix we give more details about the electric-magnetic duality invariance enjoyed by our new D-term.
We will formulate the procedure in four-dimensional $\mathcal{N}=1$ supersymmetry,
which is sufficient to illustrate how the duality works in our setup,
but the results can then be easily lifted to supergravity.

In four dimensions the dual theory of a gauge vector is again a gauge vector theory.
The electric-magnetic duality is a symmetry that operates on the level of the equations of motion of the abelian gauge vector.
It is realized by exchanging the electric field $E_i = F_{0i}$ with the magnetic field $B_i = - \frac12 \varepsilon_{ijk} F_{jk}$, and asking the equations of motion to remain invariant.
Therefore it is a property that holds on-shell, namely when we are using the equations of motion.
On the manifest Lorentz covariant formulation the duality acts by exchanging the equations of motion for the field-strength of the gauge vector with the Bianchi identity for the dual field-strength and vice versa,
in other words the duality acts by exchanging the field-strength $F_{mn}$ by the dual field-strength $\varepsilon^{k\ell mn} F_{mn}$.

The formulation of the duality within a supersymmetric setup can be found for example in \cite{Bagger:1996wp},
which we will follow here.
In this appendix we follow the superspace conventions of \cite{Wess:1992cp}.
The supersymmetric procedure starts from a Lagrangian written in terms of the vector multiplet's $V$ superfield field-strength
\be
\label{WBI}
W_\alpha = - \frac14 \overline D^2 D_\alpha V \, ,
\ee
which satisfies the supersymmetric embedding of the Bianchi identity $\partial_{[m} F_{k\ell]}=0$
in the form $D^\alpha W_\alpha = \overline D_{\dot \alpha} \overline W^{\dot \alpha}$.
Subsequently one relaxes the Bianchi identity,
which stems from \eqref{WBI} and introduces the term
\be
\label{B2}
\frac{\rmi}{2} \int {\rm d}^2 \theta\, Z^\alpha W_\alpha + \hc \supset F_{k\ell} \, \varepsilon^{k\ell mn} \, B_{mn} \, .
\ee
Here $W_\alpha$ is unconstrained, but we have introduced the field-strength superfield of the dual vector multiplet $U$ ($U$ is a real superfield),
namely
\be
\label{ZU}
Z_\alpha = - \frac14 \overline D^2 D_\alpha U \, ,
\ee
which contains the field-strength for the dual vector $B_{mn} = \partial_m C_n - \partial_n C_m$,
where
\be
C_m = \frac14 [ D_\alpha , \overline D_{\dot \alpha} ] U | \, .
\ee
By integrating out $U$ we get \eqref{WBI} and the theory takes the original form.
However by integrating out $W_\alpha$ we get the dual theory in terms of $Z_\alpha$.
Essentially one has to show that the theory takes the same form when $U$ is integrated out
with the form it has when $W_\alpha$ is integrated out.
In this way the field-strength of the gauge vector is exchanged with the dual field-strength,
within a setup that manifestly preserves supersymmetry.

We will now turn to the theory
\be
\label{B10}
{\cal L}_\text{NEW} = \frac14 \left( \int d^2 \theta \, W^2(V) + \text{h.c.} \right)
+ 8 \int d^4 \theta \, \xi \frac{W^2 \overline W^2 }{D^2 W^2  \overline D^2 \overline W^2 } D^\alpha W_\alpha
\ee
and show how it preserves the aforementioned duality.
Since the duality operates on the level of the equations of motion we are allowed to use any form of the theory,
as long as it is on-shell equivalent to the original model \eqref{B10}.
We will therefore assume one follows the procedure presented in section \ref{sec:d_term_nilpotent} of the article,
and in this way the theory takes the form
\begin{equation}
\label{B11}
{\cal L} = \int d^4 \theta\, X \overline X +  \frac14 \left( \int d^2 \theta \, \tilde W^2(V) + \text{h.c.} \right)
- \frac{\xi}{\sqrt 2} \left( \int d^2 \theta \,  X + \text{h.c.} \right) \, .
\end{equation}
The superfields satisfy
\be
\label{cB1}
X^2 = 0 \  ,\qquad  \ X \, \tilde W_\alpha = 0 \ ,\qquad
 \ |X|^2 D^\alpha \tilde W_\alpha = 0 \, .
\ee
We will show that the gauge field belonging to the vector multiplet $\tilde V$ in the Lagrangian \eqref{B11}  enjoys the electric-magnetic duality.

To prove the duality we will bring the theory in a simpler form.
In particular we will show that the last constraint 
in \eqref{cB1} can be relaxed on-shell.
This observation makes the electric-magnetic duality much easier to prove in the supersymmetric setup, following the method of  \cite{Bagger:1996wp}.

We impose the middle equation in (\ref{cB1}) with a chiral spinor Lagrange multiplier field $B_\alpha $. Including also  (\ref{B2}), we write
\begin{align}
  {\cal L} & = \int d^4 \theta\, X \overline X +  \frac14 \left( \int d^2 \theta \, \tilde W^2 + \text{h.c.} \right)
- \frac{\xi}{\sqrt 2} \left( \int d^2 \theta \,  X + \text{h.c.} \right)  \nonumber\\
    & +\left(-\frac{\rmi}{8} \int {\rm d}^2 \theta\, (\overline D^2 D^\alpha \tilde U ) \tilde W_\alpha +\frac{1}{2} \int d^2 \theta\, B^\alpha \tilde W_\alpha X + \hc\right)\,,
\label{Lcomplete}
\end{align}
where $X$ is nilpotent ($X^2=0$), $\tilde W_\alpha $, $\tilde U$  and $B_\alpha $ are unconstrained. First we prove the equivalence with the Lagrangian (\ref{B11}) with the constraints (\ref{cB1}). 
By varying $\tilde U$ we find that
\begin{equation}
  \tilde W_\alpha = - \frac14 \overline D^2 D_\alpha \tilde V \, ,
 \label{solnWinV}
\end{equation}
where $\tilde V$ is real but otherwise unconstrained. Varying $\tilde V$ and $B_\alpha $ we find
\be
\label{W2}
\delta \tilde V \ \  &:& \  \  D^\alpha  \tilde W_\alpha  + D^\alpha (X B_\alpha) + \overline D_{\dot \alpha} (\overline X \, \overline B^{\dot \alpha}) = 0  \, ,
\\
\label{BXW2}
\delta B_\alpha \ \  &:& \  \ X \tilde W_\alpha =0 \, .
\ee
We can multiply  (\ref{W2}) with $|X|^2$, using (\ref{BXW2}) to derive also the last of (\ref{cB1}).
Therefore we have proved that the latter constraint, even if it is not imposed by a Lagrange multiplier,
will emerge from the equations of motion, and we can relax it in any situation where we want to study an equivalent theory.

Instead of varying $\tilde U$ in (\ref{Lcomplete}), we now vary $\tilde W_\alpha$ and $B_\alpha$ to get
\be
\label{WWW}
\delta \tilde W_\alpha \ \  &: &\  \  \tilde W_\alpha - \frac{\rmi}{4} \overline D^2 D_\alpha \tilde U + X B_\alpha = 0 \ ,
\\
\label{BBB}
\delta B_\alpha \ \  &:& \  \ X \tilde W_\alpha = 0  \, .
\ee
From here we see that by replacing $\tilde W_\alpha$ back into \eqref{Lcomplete} and using \eqref{WWW} and \eqref{BBB} we get
\be
{\cal L}_\text{dual} = \int d^4 \theta X \overline X +  \frac14 \left( \int d^2 \theta \, \tilde Z^2(\tilde U) + \text{h.c.} \right)
- \frac{\xi}{\sqrt 2} \left( \int d^2 \theta \,  X + \text{h.c.} \right) \, ,
\ee
where now $\tilde Z_\alpha$ is the field-strength superfield of the constrained vector multiplet $\tilde U$ which satisfies 
\be
\tilde Z_\alpha =- \frac14 \overline D^2 D_\alpha \tilde U \,,\qquad X \tilde Z_\alpha = 0 \, .
\ee

Now following the inverse procedure of the one described in appendix A and in section \ref{sec:d_term_nilpotent},
one can go from the non-linear realization of the theory in terms of constrained superfields $X$ and $\tilde Z_\alpha(\tilde U)$,
back to the linear realization in terms of \eqref{ZU} and find
\be
\label{B16}
{\cal L}_\text{NEW-dual} = \frac14 \left( \int d^2 \theta \, Z^2(U) + \text{h.c.} \right)
+ 8 \int d^4 \theta \, \xi \frac{Z^2 \overline Z^2 }{D^2 Z^2  \overline D^2 \overline Z^2 } D^\alpha Z_\alpha \, .
\ee
This completes the proof of the equivalence between the theory \eqref{B10} and the theory \eqref{B16}.
However, when \eqref{B10} is describing the interactions of the electric field, then \eqref{B16} will describe the interactions of the magnetic,
and vice versa,
and since they are equivalent,
the theory enjoys electric-magnetic duality.

%%%%%%%%%%%%%%%%%%%%%%%%%%%%%%%%%%%%%%%%%%%%%%%%%%%%%%%%%%%%%%

%%%%%%%%%%%%%%%%%%%%%%%%%%%%%%%%%%%%%%%%%%%%%%%%%%%%%%%%%
%\bibliography{supergravity}
%%Included for WinEdt Gather Purpose (do not remove the comment line below:
%             %input "C:\localtexmf\bibtex\bib\*.bib"
%             %input "C:\Program Files\MiKTeX\texmf\bibtex\bib\*.bib"
%\bibliographystyle{toine}

\providecommand{\href}[2]{#2}\begingroup\raggedright\endgroup

\end{document}